\documentclass[prb,twocolumn,a4paper,showpacs,superscriptaddress,floatfix]{revtex4} 

\usepackage{amsmath}
\usepackage{bm}
\usepackage{amssymb}

\usepackage{graphicx}

\begin{document}
\draft

\title{Magnetoplasmons bound to short-range impurities in graphene: \\ Symmetries and optics}

\author{Andrea M.\ Fischer}
\affiliation{Institute of Advanced Study and Department of Physics,
  University of Warwick, Coventry CV4 7AL, United Kingdom}
\author{Rudolf A.\ R\"omer}
\affiliation{Department of Physics and Centre for Scientific Computing,
  University of Warwick, Coventry CV4 7AL, United Kingdom}
\author{Alexander B.\ Dzyubenko}
\affiliation{Department of Physics, California State University Bakersfield, Bakersfield, CA 93311, USA}
\affiliation{General Physics Institute, Russian Academy of Sciences, Moscow 119991, Russia}

\date{$Revision: 1.22 $, compiled \today}

\begin{abstract}
We consider a graphene sheet in the presence of a strong perpendicular magnetic field with a single short-range $\delta$-impurity situated at one of the carbon sites. We study the neutral inter-Landau level collective excitations, magnetoplasmons, which become localized on the impurity. 
Some of these excitations involve a pseudospin flip (intervalley transitions), 
since the impurity can scatter electrons between the two valleys. 
We propose a classification of states of the excitations in graphene and introduce the appropriate quantum numbers.
The energies and optical strengths of collective excitations are calculated for a range of integer filling factors and impurity strengths. 
We establish a set of symmetries matching the energies and absorption strengths of collective excitations for different sublattice locations of the impurity, filling factors, circular light polarizations and signs of the impurity potential.
\end{abstract}


\maketitle

\section{Introduction}
\label{sec-introduction}
At first glance, the electronic structure of graphene is simple: due to the $sp^2$ bonding of the single-atom-thick two-dimensional (2D) carbon layer, it is easily modeled by a non-interacting tight-binding approach.\cite{Wal47,Sem84} For the 2D honeycomb lattice, this results  at low energies in two non-equivalent Dirac cones at the $\mathbf{K}$ and $\mathbf{K}'$ points. 
And indeed much of the tremendous progress\cite{CasGPN09,AbeABZ10} in understanding graphene's electronic properties\cite{KosVSL11} since its isolation\cite{NovGMJ04} in 2004 has been due to the exploitation of this fact. 
However, electron-electron ($e$-$e$) interactions\cite{KotUPC10,NomM06} also play an important role in graphene, particularly in the high magnetic field regime.\cite{NovGMJ05,ZhaTSK05,DuSDL09,BolGSS09}
As well as the transport properties, $e$-$e$ interactions affect the optical excitations of the system. Previous studies have calculated the dispersion relation for particle-hole excitations for the case of clean graphene with integer Landau level 
(LL) filling.\cite{IyeWFB07,BycM08,RolFG10b} This has been done for the bilayer system too.\cite{TahS08,Shi09} Furthermore, charged collective excitations have been predicted to exist as discrete states outside the continuum.\cite{FisRD10} 

In this work, we investigate the effect of \emph{short-ranged} disorder on the collective excitations (CEs) of the monolayer system in the presence of a strong perpendicular magnetic field. Specifically, we calculate the energies and optical properties of CEs that become localized on a $\delta$-function scatterer situated at one of the graphene lattice sites. We shall refer to this scatterer as an impurity, but it can represent a vacancy as well.\cite{PerGLP06,Bas08} We explore how the bound states are influenced by the filling factor $\nu$, the sublattice containing the impurity, the light polarization and the impurity strength. The results are notably different from those for a long-range Coulomb impurity potential, studied previously.\cite{FisDR09a} In particular, the choice of a sublattice position for the impurity breaks the equivalence between $\mathbf{K}$ and $\mathbf{K}'$ points and thus the SU(4) symmetry. Furthermore, the short-range character of the impurity allows for significant inter-valley scattering.
Nevertheless, we will show that due to particle-hole symmetry at the Dirac point, the resulting optical excitations obey another form of sublattice symmetry between attractive and repulsive impurities. Our results show a symmetry structure which allows us to distinguish between long- and short-ranged impurities in the optical excitations of graphene.
\section{Theoretical approach}
\label{sec-model}
\subsection{Single particle problem}
\label{subsec-spp}
\begin{figure}[tb]
 \centering
 \includegraphics[width=0.47\textwidth]{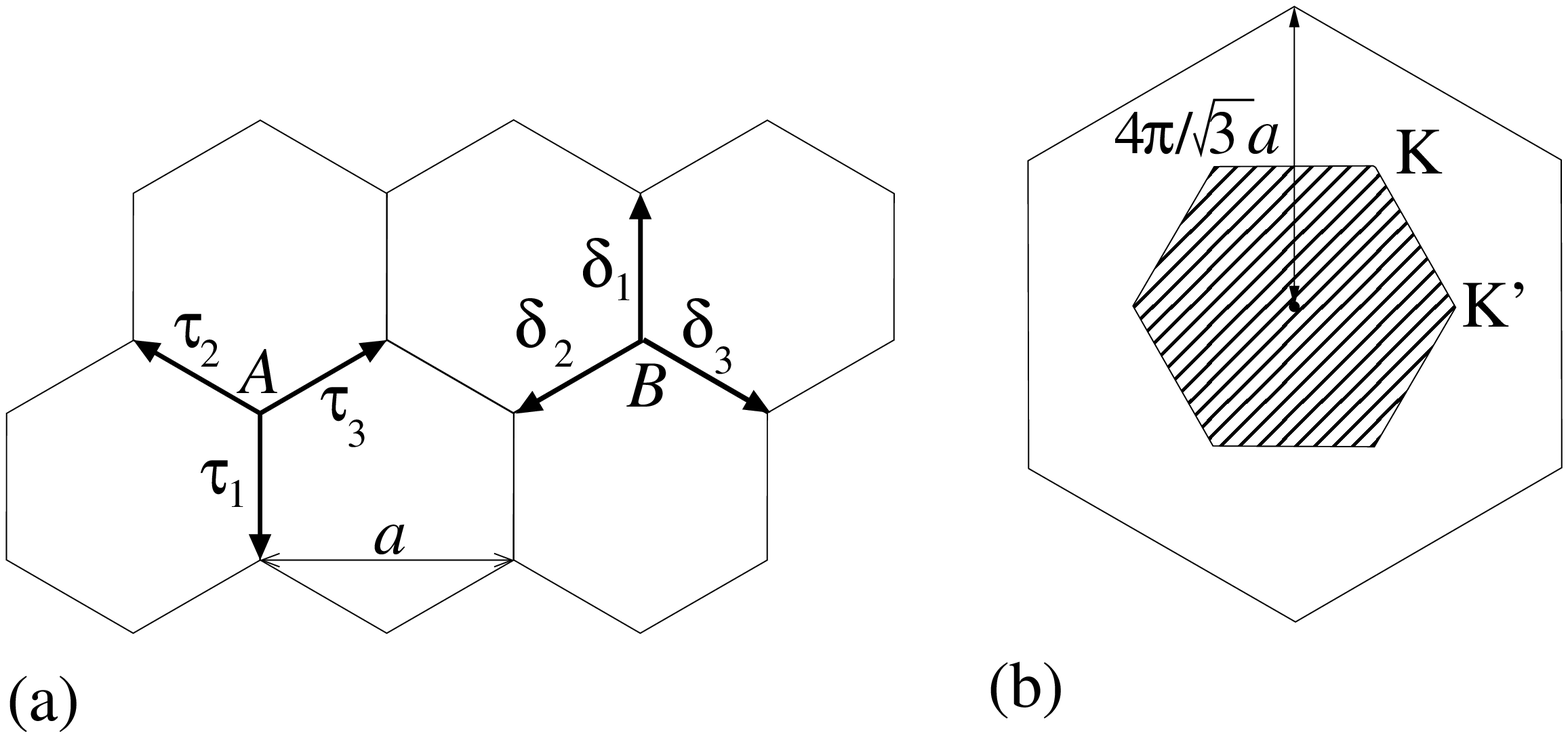}
  \caption {(a) Real space graphene lattice with nearest neighbor vectors defined for the two cases of having an $A$ or $B$ sublattice site at the origin. (b) Reciprocal lattice with the first Brillouin zone indicated by shading and definitions of $\mathbf{K}$, $\mathbf{K}'$ points.
}
\label{fig-lat}
\end{figure}

Let us consider a single electron in graphene in a perpendicular magnetic field  with a $\delta$-function impurity located at the origin. For the case when the origin is at a $\mu = A, B$ sublattice site, the Hamiltonian has the form $\hat{H}_{\mu}  =  \hat{H}_0 + \hat{V}_{\mu}$, where $\hat{H}_0$ is the Hamiltonian for a free electron in graphene in a uniform magnetic field\cite{CasGPN09}
\begin{equation}
\label{eq-ham_free}
 \hat{H}_0=v_\mathrm{F}
\left(
\begin{array}{cccc}
0  & \Pi_- & 0 & 0 \\
\Pi_+ & 0 & 0 &  0 \\
0 & 0 & 0 & \Pi_+ \\
0 &  0  & \Pi_- & 0 
\end{array}
\right).
\end{equation}
Here $v_\mathrm{F}$ is the Fermi velocity and $\Pi_\pm=\Pi_x \pm i \Pi_y$, 
with $\bm{\Pi} = \mathbf{p} + \frac{e}{c}\mathbf{A}$, the kinematic momentum operator. 
Note that the $(A\mathbf{K}, B\mathbf{K}, A\mathbf{K}', B\mathbf{K}')$ ordering is used and that 
$\hat{H}_0$ is diagonal with respect to the valley index. The contribution to the Hamiltonian 
from the short-range $V(\mathbf{r})=V \delta \left(\mathbf{r}\right)$ impurity is
\begin{equation}
\label{eq-va}
\hat{V}_A=
V \delta \left(\mathbf{r}\right)
\left(
\begin{array}{cccc}
1 & 0 & 1 & 0 \\
0 & 0 & 0 & 0 \\
1 & 0 & 1 & 0 \\
0 & 0 & 0 & 0
\end{array}
\right),
\end{equation}
for an impurity located on an $A$ site at the origin and
\begin{equation}
\label{eq-vb}
\hat{V}_B=
V \delta \left(\mathbf{r}\right)
\left(
\begin{array}{cccc}
0 & 0 & 0 & 0 \\
0 & 1 & 0 & -\alpha^\ast \\
0 & 0 & 0 & 0 \\
0 & -\alpha & 0 & 1 
\end{array}
\right),
\end{equation}
for an impurity located on a $B$ site at the origin. 
Here $\alpha=e^{2\pi i/3}$ and $V=\sqrt{3}Wa^2/2$, where $W$ is the onsite energy associated with the impurity, 
see Eq.~(\ref{eq-tbhama}); $a$ is the distance between atoms on the \emph{same} sublattice, see Fig.~\ref{fig-lat}. 
The off-diagonal terms in Eqs.~(\ref{eq-va}) and (\ref{eq-vb})
describe scattering between the valleys; such terms can be neglected for long-range potentials.
Note that in both cases $\mu = A, B$ the origin is chosen at the impurity. The derivation 
of Eqs.~(\ref{eq-ham_free})--(\ref{eq-vb}) is given in Appendix \ref{app-ham}; see also Ref.~\onlinecite{AndN98}.
The eigenstates of the Hamiltonian $\hat{H}_0$ in the symmetric gauge ${\bf A} = \frac12 {\bf B} \times {\bf r}$ for an electron in, e.g., 
the ${\bf K}$ valley (pseudospin $\tau = \: \,\Uparrow$),
are 
\begin{align}
\label{sp}
\Psi_{n s \Uparrow m}(\mathbf{r}) & = 
\langle \mathbf{r} |  c^{\dag}_{n s \Uparrow m} |0 \rangle \\ \nonumber
& =\Phi_{n \Uparrow m}(\mathbf{r}) \, \chi_s\\ \nonumber
                & = a_n
                     (\mathcal{S}_n \phi_{|n|- 1 \,  m}(\mathbf{r}),
                          \phi_{|n| \,  m}(\mathbf{r}),
                                     0,
                                     0)^{\mathrm{T}}  \, \chi_s \, .
\end{align}
Here $n$ is the integer LL number, $\phi_{|n|  m}({\bf r})$ is a 2D Electron Gas (2DEG) wavefunction
with oscillator quantum number $m = 0, 1, \ldots$, $a_n=2^{\frac{1}{2}(\delta_{n,0} -1)}$,
$\mathcal{S}_n={\rm sign}(n)$ (with $\mathcal{S}_0=0$) and
$\chi_s$ is the spin part for the two spin projections
$s = \: \, \uparrow, \downarrow$. The corresponding wavefunction in the ${\bf K'}$ valley ($\tau = \: \, \Downarrow$) is obtained by reversing the order of the spinor components. There is the Landau degeneracy in quantum number $m$, 
so that defining the composite index $\mathcal{N} = \{ n s \tau \}$, the single particle energy is 
\begin{equation}
  \label{eq-en}
\epsilon_\mathcal{N} = \mathcal{S}_n \hbar\omega_c \sqrt{|n|} + \hbar\omega_s s_z + \hbar\omega_{v} \tau_z   \, .
\end{equation}
Here $\hbar\omega_c = \sqrt{2} \, \hbar  v_F/ \ell_B$ is the cyclotron energy in graphene, $\ell_B=\sqrt{\hbar c/eB}$ is the magnetic length 
and $\hbar\omega_s, \hbar\omega_v$ the phenomenological spin and valley splittings, respectively.
We assume $\hbar\omega_s > \hbar\omega_v$ and that these splittings are small in comparison with the cyclotron energy,
 $\hbar\omega_s, \hbar\omega_v \ll \hbar\omega_c$;
we set them equal to zero for the purposes of numerical calculations. 
Note that the wavefunctions $\Psi_{\mathcal{N} m}(\mathbf{r})$ are not eigenfunctions of the orbital angular momentum projection operator
$\hat{l}_z = \mathbf{r} \times \mathbf{p} $. 
Instead, the generalized angular momentum\cite{DivM84}
$\hat{j}_z=\hat{l}_z+\frac{1}{2} \sigma_z \otimes \hat{I}$ is conserved in graphene. 
Here $\sigma_z$ is a Pauli matrix acting on the isospin (in the sublattice index space
$\mu = A, B$) while the unit matrix  $\hat{I}$ acts on the pseudospin 
(in the valley ${\bf K}, {\bf K'}$ space $\tau  = \: \, \Uparrow, \Downarrow$). 
The generalized angular momentum operator gives
 $\hat{j}_z \Psi_{\mathcal{N} m}(\mathbf{r})=j_z\Psi_{\mathcal{N} m}(\mathbf{r})$, 
with the half-integer eigenvalue $j_z=l_z - \frac{1}{2}$ 
and the integer orbital part  $l_z = |n|-m$.

The calculation of impurity matrix elements involving $\hat{V}_{\mu}$ is straightforward, since
\begin{align}
\label{eq-impme}
    {\mathcal{V}_{\mu}}_{\mathcal{N} m}^{\mathcal{N}' m'}& = 
       \int \! \mathrm{d} \mathbf{r} \, \Psi^\dag_{\mathcal{ N'}m'}(\mathbf{r}) 
	            \hat{V}_{\mu } \Psi^{\vphantom{\dagger}}_{ \mathcal{N}m}(\mathbf{r}) \\ \nonumber
   & = \delta_{s, s'} \int \! \mathrm{d} \mathbf{r} \, \Phi^\dag_{n' \tau' m'}(\mathbf{r}) \hat{V}_{\mu } \Phi^{\vphantom{\dagger}}_{n \tau m}(\mathbf{r})        \\ 
   & \sim \delta_{s, s'} V \, \phi_{k  m'}^*\left( 0 \right)\phi^{\vphantom{\dagger}}_{l  m}\left( 0 \right) \nonumber,
\end{align}
where $l \in \left\lbrace |n| ,|n|-1  \right\rbrace$ 
 and  $k \in \left\lbrace |n'|,|n'|-1 \right\rbrace$, 
with exact values set by pseudospins (valley indices) $\tau$ and $\tau'$. 
In addition, $\phi_{n m}\left(0 \right) \sim \delta_{n,m} $, 
imposing the selection rule determining which states are affected by the $\delta$-impurity. 
Indeed, only the $s$-orbitals with $l_z = |n| - m = 0$ have non-vanishing probability amplitudes at the origin, where the $\delta$-impurity is located. 
Note that this selection rule allows mixing of states with \emph{different} generalized angular momentum projections $j_z$.
This is in contrast to a Coulomb potential, where $j_z$ is strictly conserved. 
In Appendix \ref{app-high} we dicuss when higher order corrections to the energies $\epsilon_\mathcal{N}$ 
due to the impurity interaction are small and may be ignored. 
\begin{figure}[t]
 \centering
 \includegraphics[width=0.47\textwidth]{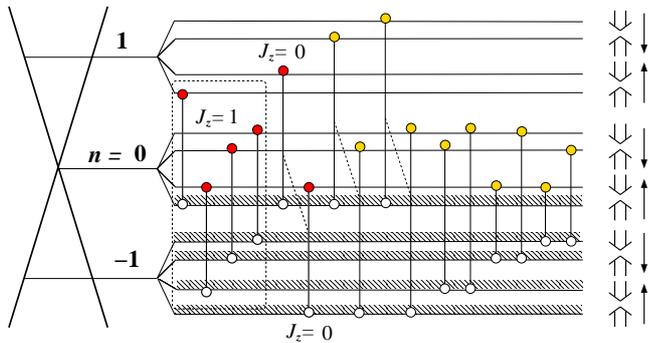}
  \caption {(Color online) Particle-hole excitations for filling factor $\nu=-1$. 
  The thin arrows indicate two possible spin states $s= \: \, \uparrow, \downarrow$
  and the thick arrows two possible pseudospin $\tau  = \: \, \Uparrow, \Downarrow$ 
  (valleys  ${\bf K}, {\bf K'}$) states. 
The dashed lines connect excitations mixed by the direct Coulomb interaction, see Fig.~\protect\ref{fig-Coul}a. 
The dashed box contains four excitations resonantly mixed by the exchange Coulomb interaction, see Fig.~\protect\ref{fig-Coul}b.
The six leftmost (red-dotted) excitations are those which are mixed when there is an impurity on the $A$ sublattice and are optically-active in the left circular polarization $\sigma^+$ of light.
}
\label{fig-diag}
\end{figure}

\subsection{Collective excitations}
\label{subsec-ces}

We consider excitations in which an electron is promoted from one of the uppermost filled LLs, $\mathcal{N}_2$ in the Dirac sea, to an empty higher lying LL $\mathcal{N}_1$, leaving behind a hole.
The creation operator for such an excitation is
\begin{equation}
\label{eq-Qdag} 
 \mbox{} \hspace{-5pt}
      Q^{\dag}_{\mathcal{ N}_1  \mathcal{ N}_2 J_z }  =
          \sum_{m_1 , m_2 = 0}^{\infty}
           A_{\mathcal{N}_1 \mathcal{N}_2 J_z }(m_1,m_2) \,
           c^{\dag}_{\mathcal{N}_1 m_1} d^{\dag}_{\mathcal{N}_2 m_2} \, ,
\end{equation}
where the hole representation, $c_{\mathcal{N} m} \rightarrow d^{\dag}_{\mathcal{N} m}$
and $c^{\dag}_{\mathcal{N} m} \rightarrow d_{\mathcal{N} m}$, is used for all filled levels. 
The excitation operator $Q^{\dag}_{\mathcal{ N}_1  \mathcal{ N}_2 J_z }$ acts on the ground state, 
denoted by $|\nu \rangle$ for a system with filling factor $\nu$.
The expansion coefficients satisfy 
${A_{\mathcal{N}_1 \mathcal{N}_2 J_z }(m_1,m_2) \sim \delta_{J_z, |n_1| - m_1 - |n_2| + m_2}}$.
The quantum number $J_z$ is the total generalized angular momentum projection given by summing over the $j_z$ values for individual particles $i$,
$j_{zi} = l_{zi} - \frac{1}{2}$, with $l_{zi} = |n_i| - m_i$.
In second quantized form,  
\begin{equation}
\label{eq-jzop}
\hat{J_z}  =   \sum_{\mathcal{N}_1, m_1} j_{z1} \, c^{\dag}_{\mathcal{N}_1 m_1}c^{\vphantom{\dagger}}_{\mathcal{N}_1 m_1} 
             - \sum_{\mathcal{N}_2, m_2} j_{z2} \, d^{\dag}_{\mathcal{N}_2 m_2}d^{\vphantom{\dagger}}_{\mathcal{N}_2 m_2}   \, .
\end{equation}
We have 
$\hat{J_z}Q^{\dag}_{\mathcal{ N}_1  \mathcal{ N}_2 J_z }|\nu\rangle=J_zQ^{\dag}_{\mathcal{ N}_1  \mathcal{ N}_2 J_z }|\nu\rangle$. 
Note that for the neutral particle-hole excitations 
$J_z = l_{z1} - l_{z2} = |n_1| - m_1 - |n_2| + m_2$ only contains an orbital part, in contrast to the single particle states and charged
collective excitations\cite{FisRD10} in graphene.

\begin{figure}[bht]
 \centering
 \includegraphics[width=0.25\textwidth]{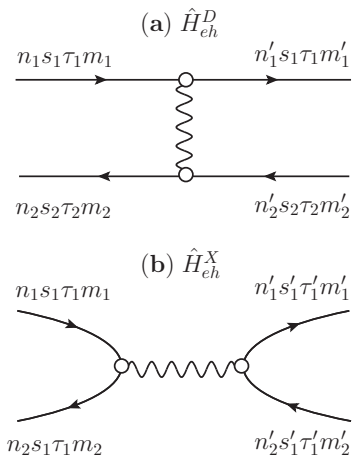}
  \caption {Direct (a) and exchange (b) electron-hole interaction vertices, see Eq.~(\protect\ref{eq-w}).
}
\label{fig-Coul}
\end{figure}

The total Hamiltonian, $\hat{H}_{\mu} =\hat{H}_{e\mu}+\hat{H}_{h\mu}+\hat{H}_{\mathrm{int}}$, 
including the free energies, the interaction with the impurity on the $\mu = A, B$ site,
and the electron-hole ($e$-$h$) interaction is
\begin{align}
\label{eq-bigham}
\hat{H}_{\mu} &= 
                 \sum_{\substack{\mathcal{N}_1, \mathcal{N}_1' \\ m_1, m_1'}}
                 \left(  \delta_{\mathcal{N}_1 \mathcal{N}_1'}\delta_{m_1 m_1'}
			           \tilde{\epsilon}_{\mathcal{N}_1}+{\mathcal{V}_{\mu}}_{\mathcal{N}_1 m_1}^{\mathcal{N}_1' m_1'} \right) 
                c^{\dag}_{\mathcal{N}_1' m_1'}c^{\vphantom{\dagger}}_{\mathcal{N}_1 m_1} \\ \nonumber
& - 
    \sum_{\substack{\mathcal{N}_2, \mathcal{N}_2' \\ m_2, m_2'}}
	\left(\delta_{\mathcal{N}_2 \mathcal{N}_2'} \delta_{m_2 m_2'}
	\tilde{\epsilon}_{\mathcal{N}_2}+{\mathcal{V}_{\mu}}_{\mathcal{N}_2 m_2}^{\mathcal{N}_2' m_2'} \right)  
	d^{\dag}_{\mathcal{N}_2' m_2'} d^{\vphantom{\dagger}}_{\mathcal{N}_2 m_2} \\ \nonumber
& + \sum_{\substack{\mathcal{N}^{\vphantom{A}}_1,  \mathcal{N}^{\vphantom{A}}_2  \\  m^{\vphantom{a}}_1, m_2 } }
    \sum_{\substack{\mathcal{N}_1', \mathcal{N}_2' \\ m_1', m_2'} }
	     \mathcal{W}_{\mathcal{N}_1 m_1   \mathcal{N}_2 m_2}^{\mathcal{N}_1' m_1'  \mathcal{N}_2' m_2'} \, 
         c^{\dag}_{\mathcal{N}_1' m_1'}d^{\dag}_{\mathcal{N}_2' m_2'}
		 d^{\vphantom{\dagger}}_{\mathcal{N}_2 m_2}c^{\vphantom{\dagger}}_{\mathcal{N}_1 m_1}. \nonumber  \,
\end{align}
The $\tilde{\epsilon}_\mathcal{N}$ are the single particle energies, which are renormalized 
due to the exchange interaction with other electrons in the Dirac sea (see Appendix \ref{app-se}). 
The last term gives the dynamical part of the $e$-$h$ Coulomb interaction, following from the pairwise Coulomb potential 
of $e$-$e$ interactions in graphene,
$U\left( |\mathbf{r}_1-\mathbf{r}_2|\right) = \frac{e^2}{\varepsilon |\mathbf{r}_1-\mathbf{r}_2|}$.
Here $\varepsilon$ is an effective dielectric constant,
which depends for graphene on its environment. 
The dynamical $e$-$h$ interaction is made up of the 
$e$-$h$ direct attraction and the $e$-$h$ exchange repulsion
$\hat{H}_{\mathrm{int}} = \hat{H}_{eh}^{D} + \hat{H}_{eh}^{X}$ with the total vertex given by 
\begin{equation}
\label{eq-w} 
 \mathcal{W}_{\mathcal{N}_1 m_1   \mathcal{N}_2  m_2 }^{\mathcal{N}_1' m_1'  \mathcal{N}_2' m_2'}=
-\mathcal{U}_{\mathcal{N}_1 m_1   \mathcal{N}_2' m_2'}^{\mathcal{N}_1' m_1'  \mathcal{N}_2  m_2 }
+\mathcal{U}_{\mathcal{N}_1 m_1   \mathcal{N}_2' m_2'}^{\mathcal{N}_2  m_2   \mathcal{N}_1' m_1'} \, ,
\end{equation}
see Fig.~\ref{fig-Coul}. 
The matrix element $\mathcal{U}$ is defined in the \emph{electron} representation by
\begin{widetext}
\begin{equation}
\label{eq-u} 
   \mathcal{U}_{\mathcal{N}_1 m_1   \mathcal{N}_2 m_2}^{\mathcal{N}_1' m_1'  \mathcal{N}_2' m_2'} = 
   \delta_{s_1, s_1'}\delta_{s_2, s_2'} \int \!\! \mathrm{d} \mathbf{r}_1 \!\! \int \!\! \mathrm{d} \mathbf{r}_2 \; 
    \Phi^\dag_{ n_1'\tau_1' m_1'}(\mathbf{r}_1)\otimes\Phi^\dag_{ n_2'\tau_2'm_2'}(\mathbf{r}_2)
       U\left( |\mathbf{r}_1-\mathbf{r}_2|\right)
	\Phi^{\vphantom{\dagger}}_{ n_1 \tau_1 m_1} (\mathbf{r}_1) \otimes \Phi^{\vphantom{\dagger}}_{ n_2 \tau_2 m_2 } (\mathbf{r}_2) \, , 
\end{equation}
\end{widetext}
where $\otimes$ denotes the direct (Kronecker) product.
Note that $\mathcal{U}$ may be expressed in terms of the corresponding matrix elements for the 2DEG\cite{FisDR09a} 
and conserves spin, pseudopsin (no intervalley scattering), and the generalized angular momentum:
\begin{equation}
\label{eq-udeltas}
 \mathcal{U}_{\mathcal{N}_1 m_1   \mathcal{N}_2 m_2}^{\mathcal{N}_1' m_1'  \mathcal{N}_2' m_2'} 
 \sim \delta_{s_1, s_1'}\delta_{s_2, s_2'}\delta_{\tau_1, \tau_1'}\delta_{\tau_2, \tau_2'} \delta_{J_z, J_z'}. \,
\end{equation}
It also possesses the SU(4) symmetry,\cite{AbeABZ10,GoeMD06,FisRD10} as described below.

\subsection{Symmetry properties of the Hamiltonian}
\label{subsec-hamsym}

In the absence of an impurity ($V=0$), the Hamiltonian $\hat{H} = \hat{H}_0 + \hat{H}_{\mathrm{int}}$
has several symmetries. 
Firstly, there is the generalized axial symmetry\cite{DivM84,FisDR09a}
$[\hat{J_z},\hat{H}]=0$. 
This symmetry is broken in the presence of a $\delta$-impurity, which scatters particles between the valleys.
Specifically for $V \ne 0$,
\begin{eqnarray}
    \label{eq-jzcom}
       [\hat{J_z},\hat{H}_{\mu}]  
	& = & 
	    \sum_{\substack{\mathcal{N}_1, \mathcal{N}_1' \\ m_1, m_1'} }
		   {\mathcal{V}_{\mu}}_{\mathcal{N}_1 m_1}^{\mathcal{N}_1' m_1'} \, 
            \left( j_{z1}' - j_{z1} \right)  c^{\dag}_{\mathcal{N}_1' m_1'}c^{\vphantom{\dagger}}_{\mathcal{N}_1 m_1} \\ 
		\nonumber
    & +&
	    \sum_{\substack{\mathcal{N}_2, \mathcal{N}_2' \\ m_2, m_2'} }
		   {\mathcal{V}_{\mu}}_{\mathcal{N}_2 m_2}^{\mathcal{N}_2' m_2'} \, 
           \left( j_{z2}' - j_{z2} \right) \, d^{\dag}_{\mathcal{N}_2' m_2'}d^{\vphantom{\dagger}}_{\mathcal{N}_2 m_2} \neq 0 \, ,
\end{eqnarray}
with $j_{zi}' - j_{zi} = |n_i'|-m_i'-|n_i|+m_i$. 
This results in excitations with different $J_z$ quantum numbers being mixed by the impurity,
which we shall discuss in Sec.\ \ref{subsec-mix}. Therefore, $J_z$ ceases to be an exact quantum number.
We will still use it for classification of states to indicate their origin and predominant character.

Another kind of symmetry present in clean graphene with long-range Coulomb interactions is the SU(4) symmetry, 
due to the equivalence of the four possible pseudospin-spin states, 
$\Uparrow\uparrow$,$\Downarrow\uparrow$,$\Uparrow\downarrow$,$\Downarrow\downarrow$. 
The generators of the SU(4) group may be expressed in our context in second quantized form as 
$C_{jk} =\sum_N c^{\dagger}_{Nj} c_{Nk} - \sum_{N}d^{\dagger}_{Nk} d_{Nj}$, 
where $N=\{n m\}$ is a set of \emph{orbital} quantum numbers 
and $j,k= \, \Uparrow\uparrow$,$\Downarrow\uparrow$,$\Uparrow\downarrow$,$\Downarrow\downarrow$ are the four SU(4) \emph{flavors}. 
In the absence of an impurity and for a fully filled LL, $[\hat{H},C_{jk}]=0$. 
Upon introducing an impurity, the commutation relation becomes
\begin{eqnarray}
\label{eq-sufourcom}
[\hat{H}_{\mu},C_{jk}] & = & \sum_{N_1,N_1'} \sum_{f} \left(1+\delta_{N_1, N_1'}\right)  \\ \nonumber
                 &   & \times \left( {\mathcal{V}_{\mu}}_{N_1 j}^{N_1' f} \,  c^{\dagger}_{N_1' f} c^{\vphantom{\dagger}}_{N^{\vphantom{.}}_1 k} 
				                   - {\mathcal{V}_{\mu}}_{N_1 f}^{N_1' k} \,  c^{\dagger}_{N_1' j} c^{\vphantom{\dagger}}_{N^{\vphantom{.}}_1 f}
								    \right) \\ \nonumber
                 &   & + \sum_{N_2,N_2'} \sum_{f} \left( 1 + \delta_{N_2, N_2'} \right) \\ \nonumber
                 &   &   \times \left( {\mathcal{V}_{\mu}}_{N_2 k}^{N_2' f} \, d^{\dagger}_{N_2' f}d^{\vphantom{\dagger}}_{N^{\vphantom{.}}_2 j}
				                      -{\mathcal{V}_{\mu}}_{N_2 f}^{N_2' j} \, d^{\dagger}_{N_2' k}d^{\vphantom{\dagger}}_{N^{\vphantom{.}}_2 f}
									\right).
\end{eqnarray}
For an impurity which does not change the flavor of a particle \emph{and} has scattering matrix elements which are flavor independent,
${\mathcal{V}_{\mu}}_{N k}^{N' f} = \delta_{k,f} {\mathcal{V}_{\mu}}_{N}^{N'}$, 
one can see from Eq.~(\ref{eq-sufourcom}) that the SU(4) symmetry is still preserved, $[\hat{H}_{\mu},C_{jk}] = 0$. 
This is the case for, e.g., a Coulomb impurity and explains the degeneracies of the corresponding states.\cite{FisRD11} 
In contrast, the $\delta$-impurity considered here may scatter between the valleys, 
thus changing a particle's flavor and breaking the SU(4) symmetry.

\subsection{Optical selection rules}
\label{subsec-op}

We work in the electric dipole approximation, ignoring the magnetic field component of the electromagnetic wave. 
The Hamiltonian describing the electron-photon interaction for an incoming circularly $\sigma^\pm$ polarized  beam of light is  
\begin{equation}
 \label{eq-delh}
\delta \hat{\mathcal{H}}_\pm  = \frac{i e v _\mathrm{F} \mathcal{F}}{\omega c} e^{-i\omega t}
\left(\begin{array}{cc}
\sigma_\pm  & 0  \\
0 & \sigma_\mp
\end{array}\right),
\end{equation}
where $\mathcal{F}$ represents the electric field strength, $\omega$ is the angular frequency and $\sigma_\pm =\sigma_x\pm i\sigma_y$
are the Pauli matrices acting on the isospin (sublattice $A, B$) components. 
In the dipole approximation, the linear momentum transferred to the electron by a photon is negligible; 
accordingly, no intervalley transitions can be induced. Besides, the electric field is not coupled directly to the electron spin, 
so that $\delta \hat{\mathcal{H}}_\pm$ conserves spin, i.e., no spin flips occur in electric dipole optical transitions. 
In addition, one can show that 
\[
\langle \mathcal{N}' m' | \delta \hat{\mathcal{H}}_\pm | \mathcal{N} m \rangle \sim \delta_{|n'|\mp 1,|n|}\delta_{m, m'} \, , 
\]
where $|\mathcal{N} m \rangle\equiv c^\dagger_{\mathcal{N} m}|0\rangle$. 
The single particle selection rules are thus $|n'|-|n|=\pm1$, $m=m'$, $\tau=\tau'$ and $s=s'$, 
for the $\sigma^\pm$ polarizations, where the unprimed quantum numbers describe the initial state of the electron 
and the primed quantum numbers its final state, see Refs.~\onlinecite{GusSC07,AbeF07,FisDR09a} and  references therein.

In the second quantization form and in the $e$-$h$ representation, the interaction with the circularly $\sigma^\pm$ polarized light is
\begin{equation}
\label{eq-colldelH}
  \delta \hat{\mathcal{H}}_\pm = 
  \sum_{\mathcal{N},\mathcal{N'}} 
     \langle n' | \delta \hat{\mathcal{H}}_\pm | n\rangle
   \sum_{m} c^\dagger_{\mathcal{N}' m}d^\dagger_{\mathcal{N} m} \, .
\end{equation}
Here we made use of the fact that the matrix element 
$\langle \mathcal{N}' m | \delta \hat{\mathcal{H}}_\pm | \mathcal{N} m \rangle \equiv
\delta_{s',s}\delta_{\tau',\tau}\delta_{m',m} \langle n' | \delta \hat{\mathcal{H}}_\pm |n \rangle$
is diagonal in $s, \tau, m$ and, besides, it does not depend on these quantum numbers.
The operator emerging in Eq.~(\ref{eq-colldelH}), 
\begin{equation}
\label{Q_0}
       \frac{1}{ \sqrt{N^{\vphantom{X}}_0} } \sum_{m} c^\dagger_{\mathcal{N}' m}d^\dagger_{\mathcal{N} m} 
	   =  Q^{\dag}_{\mathcal{N}'  \mathcal{N} \bm{\kappa}=0 }\ ,
\end{equation}
is proportional to the composite boson creation operator $Q^{\dag}_{\mathcal{ N}_1  \mathcal{ N}_2 \bm{\kappa}=0}$
describing a magnetoexciton 
of zero magnetic momentum $\bm{\kappa}=0$ 
with electron (hole) in the $\mathcal{N}'$ ($\mathcal{N}$) LL in graphene.
$N_0 = S/2\pi \ell_B^2$ is the LL degeneracy with $S$ the area of the graphene sheet and the operator of magnetic translations is
$\hat{\bm{\kappa}} = \sum_i (\bm{\Pi}_i - \frac{q_i}{c} \mathbf{r}_i \times \mathbf{B})$. 
Moreover, 
the electron-photon interaction is SU(4)-symmetric, $[\delta \hat{\mathcal{H}}_\pm,C_{jk}]=0$; it does not change the flavor of an electron 
and has matrix elements which are flavor independent.
This leads to the coupling of the electric dipole photon only to the flavorless boson  in the given LLs\cite{FisDR09a}
$\bar{Q}^{\dag}_{n'  n  } \equiv \sum_{s, \tau}
Q^{\dag}_{n's \tau \, n s\tau  \, \bm{\kappa}=0 }$.
In the presence of an impurity, the magnetic momentum $\bm{\kappa}$ is no longer conserved,
$[\hat{H}_{\mu},\hat{\bm{\kappa}}] \neq 0$.
In addition, a short-range impurity induces the intervalley scattering and breaks the SU(4) symmetry, further relaxing
the selection rules.

Let $d \equiv \langle \mathcal{N}_1  \mathcal{N}_2 J_z  |\delta \hat{\mathcal{H}}_\pm |\nu\rangle$ 
denote the optical dipole transition matrix element from the ground state $| \nu \rangle $ with filling factor $\nu$, to
the final state with one excitation
\begin{equation}
\label{Q} 
  Q^{\dag}_{\mathcal{ N}_1  \mathcal{ N}_2 J_z }| \nu \rangle \equiv   |\mathcal{ N}_1  \mathcal{N}_2 J_z \rangle \, . 
  \end{equation} 
The transition matrix element squared $|d|^2$ is proportional to the absorption intensity or oscillator strength for that state.
It can be shown that 
\begin{equation}
\label{sel_Q}
\langle \mathcal{N}_1  \mathcal{N}_2 J_z  |\delta \hat{\mathcal{H}}_\pm |\nu\rangle \sim \delta_{J_z, \pm 1} \, ,
\end{equation} 
so the optical selection rule for a CE is $J_z=\pm 1$.  Besides, only the excitations with the
total spin and pseudospin projections $S_z=s_{z1}-s_{z2}=0$, $T_z=\tau_{z1}-\tau_{z2}=0$ are optically active. The latter selection rule is strict.
For equal spin (pseudospin) filling of a given $n^\mathrm{th}$ LL, 
the CE states can additionally be classified by a total spin $S=0, 1$ (pseudospin $T=0,1$).
Among those states only the spin  $S=0$ and pseudospin $T=0$ singlets 
have non-zero projections onto the state $\bar{Q}^{\dag}_{n'  n  }$, which is directly coupled to photons.
Therefore, only the singlets $S=0$, $T=0$ are optically active,
while all the triplet  $S=1$ and $T=1$ states are optically dark. 
\subsection{Mixing of excitations}
\label{subsec-mix}

The long range nature of the Coulomb potential means it cannot provide a large enough change in momentum 
$|\Delta\mathbf{k}| \simeq |\mathbf{K} - \mathbf{K}'| \sim 1/a$
to scatter between the valleys. 
It also only mixes excitations with the same angular momentum projection  $J_z=|n_1|-m_1-|n_2|+m_2$. In addition, Eqs.~(\ref{eq-w}) and (\ref{eq-udeltas}) 
restrict the possible spin and pseudospin states of transitions mixed by the Coulomb interaction. 
Generally, two excitations,  $|\mathcal{N}_1  \mathcal{N}_2 J_z \rangle $ and $|\mathcal{N}_1'  \mathcal{N}_2' J_z \rangle$, 
are mixed by (i) the direct interaction if $s_1=s_1^{'}$, $\tau_1=\tau_1^{'}$, $s_2=s_2^{'}$ and $\tau_2=\tau_2^{'}$
(see Fig.~\ref{fig-Coul}a)
and (ii) the exchange interaction if $s_1=s_2$, $\tau_1=\tau_2$, $s_1^{'}=s_2^{'}$ and $\tau_1^{'}=\tau_2^{'}$ (see Fig.~\ref{fig-Coul}b). 
Figure~\ref{fig-diag} shows the possible excitations for filling factor $\nu=-1$. 
The four excitations with no spin or pseudospin flips contained within the dashed box are mixed by the exchange interaction. 
Pairs of excitations connected by a dashed line are mixed by the direct interaction. The remaining excitations are unmixed by the $e$-$h$ Coulomb interparticle interactions.

In our calculations we assume all LLs with $n<0$ are filled, all LLs with $n>0$ are empty and that the sublevels 
of the zeroth LL become successively completely filled ($\nu=-1,0,1,2$).
We have seen that infinitely many excitations $|\mathcal{ N}_1  \mathcal{N}_2 J_z \rangle$ with the same $J_z$ 
and particular spin and pseudospin quantum numbers are mixed by the $e$-$h$ Coulomb interaction. 
However, we may truncate the basis so as to obtain a tractable finite-size matrix representation 
of Hamiltonian $\hat{H}_{\mu}$, Eq.~(\ref{eq-bigham}).
To this end we only consider excitations, which are either in resonance (have same energies) or are very close to resonance.  
In particular, we are interested in the lowest energy $\sim \hbar\omega_c$ inter-LL excitations 
for the chosen filling factors.
Therefore, we need to consider the mixing of $0 \to 1$ and $-1 \to 0$ excitations. 
Neglecting the mixing of non-resonant excitations amounts to ignoring terms with amplitudes, 
which depend on the ``fine structure constant'' for graphene, $\alpha =e^2/\hbar v_F \varepsilon$. 

But even with this simplification, the Hamiltonian matrix remains infinite because of the macroscopic degeneracy of each LL 
in the oscillator quantum number $m$. However, as long as we are interested in the states localized on the impurity, we may truncate the
summation over $m$ in Eq.~(\ref{eq-Qdag}) at a finite sufficiently large value $m_{\mathrm{max}}$. This is justified because the distance
from the origin (impurity) of single-particle orbitals in LLs increases with $m$ 
as 
\begin{equation}
   \label{dist} 
\langle \mathbf{r}^2\rangle = (2|n|+2m+1+\delta_{n,0})\ell_B^2 \, .
\end{equation}
As a result, for $e$-$h$ states localized on the impurity we have a fast convergence with $m$.\cite{Dzy90,FisDR09a} 
In our numerical calculations we choose
$m_{\mathrm{max}} =30$; the total number of states involved ranges from $180$ to $240$, 
depending on $\nu$, the sublattice of the impurity and which light polarization the excitation is bright in.

We are primarily interested in optically active excitations, although these will always be mixed by the
$\delta$-impurity potential to some which are nominally dark. 
This leads to the redistribution of the oscillator strength so that a number of states instead of being strictly dark become ``faint''.
In the previous Sec.~\ref{subsec-op}, we saw that bright states had no spin or pseudospin flips 
and $J_z=\pm 1$ for the $\sigma^\pm$ polarizations. The four excitations with no spin or pseudospin flips are always mixed by the exchange 
$e$-$h$ Coulomb interaction, although how many of them are bright depends on the filling factor and light polarization. Two or four additional excitations with pseudospin flips will be mixed to these by the $\delta$-impurity; generally, for a given $J_z$,
the admixtures will have $J_z$ and $J_z \pm 1$ values. 
Thus the full CE comprises of six or eight excitations with different spin/pseudospin characters and $J_z$ values. 
As an example, Fig.~\ref{fig-diag} indicates the six (red) excitations which should be taken into account 
for $\nu=-1$, the $\sigma^+$ light polarization and an impurity on the $A$ sublattice. 

\begin{figure}[t!]
\centering
\includegraphics[width=0.47\textwidth]{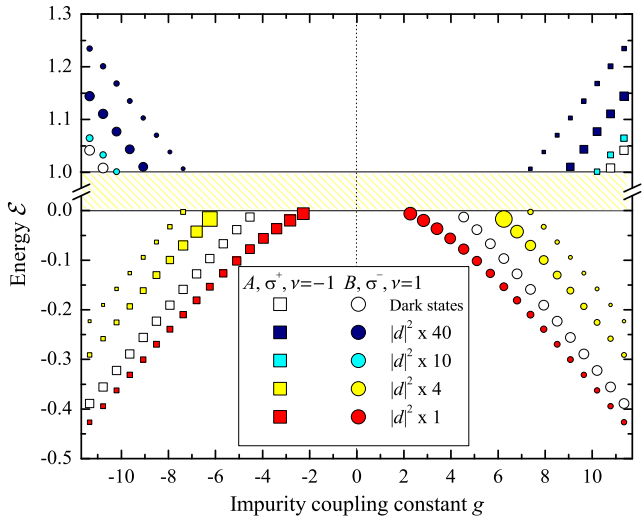}
\caption {(Color online) Dimensionless energies $\mathcal{E}$ of localized CEs 
vs dimensionless impurity coupling constant  $g$ [Eqs.~(\protect\ref{eq-ener}), (\ref{eq-impcoupconst})], 
for the ($A$, $\sigma^+$, $\nu=-1$) system (squares) and the conjugate system ($B$, $\sigma^-$, $\nu=1$) system (circles).
Notice the symmetry of the spectra, Eq.~(\protect\ref{symmetry}).
The shaded region corresponds to the magnetoplasmon band with extended states.}
\label{fig-Amz1nu1}
\end{figure}

\section{Results and discussion}
\label{sec-results}

\subsection{Energies and optical strengths}
\label{subsec-en}

\begin{figure}
  \centering
 \includegraphics[width=0.47\textwidth]{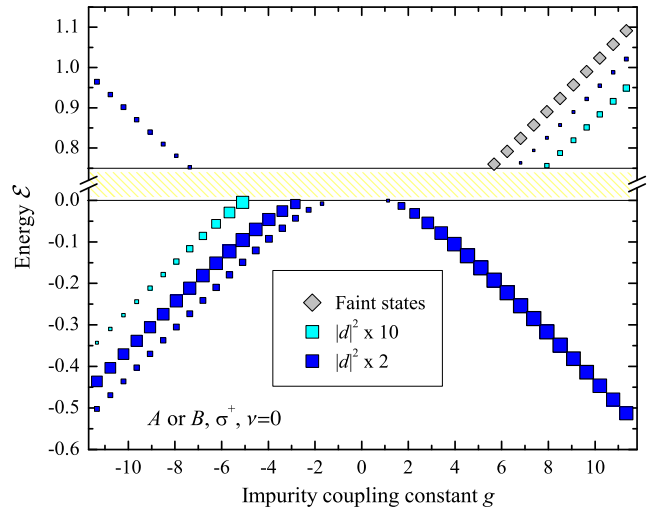}
  \caption{(Color online) Energies of localized CEs vs impurity strength as in Fig.~\ref{fig-Amz1nu1}, but for the symmetric systems ($A$, $\sigma^+$, $\nu=0$) and ($B$, $\sigma^+$, $\nu=0$). States marked as ``faint'' have values of $|d|^2$ at least three orders of magnitude smaller than the brightest states.}
\label{fig-Amz1nu2}
\end{figure}

\begin{figure}
  \centering
\includegraphics[width=0.47\textwidth]{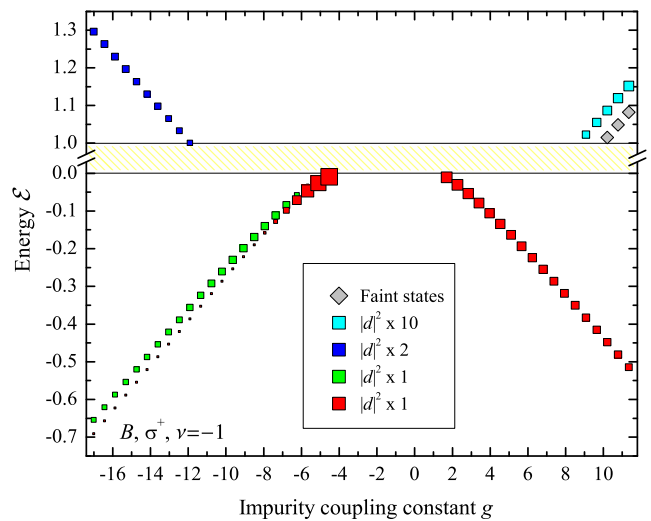}
  \caption {(Color online) Energies of localized CEs vs impurity strength as in Fig.~\ref{fig-Amz1nu2}, but
for the ($B$, $\sigma^+$, $\nu=-1$) system.
\label{fig-Bmz1nu1}}
\end{figure}

\begin{figure}[!t]
  \centering
 \includegraphics[width=0.47\textwidth]{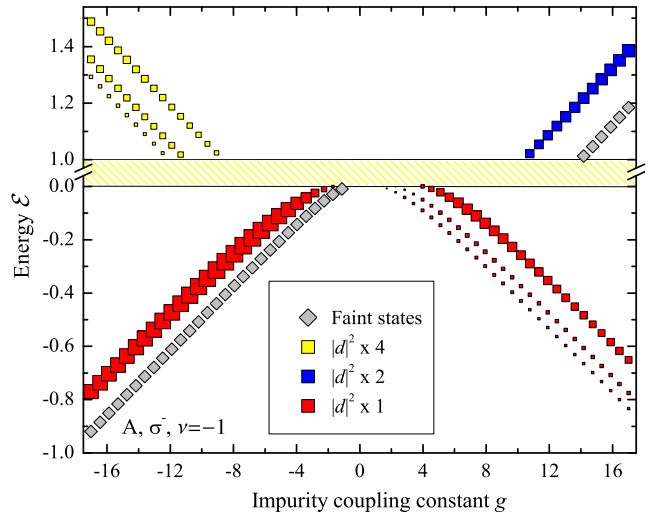}
  \caption {(Color online) Energies of localized CEs vs impurity strength as in Fig.~\ref{fig-Amz1nu2}, but
for the ($A$, $\sigma^-$, $\nu=-1$) system.
}
\label{fig-Amz-1nu1}
\end{figure}

A representative selection of results are displayed in Figs.~\ref{fig-Amz1nu1}--\ref{fig-Amz-1nu1}. 
In the absence of an impurity,\cite{BycM08,IyeWFB07,FisRD11} all CEs have \emph{extended} wavefunctions and fill a band 
(indicated by the shaded regions) of width determined by the characteristic Coulomb energy
\begin{equation}
\label{E0}
  E_0 = \sqrt{\dfrac{\pi}{2}}\dfrac{e^2}{\varepsilon\ell_B} \sim \sqrt{B} \, .
\end{equation} 
More precisely, the band has width $E_0$ for odd filling factors $\nu = -1, 1$ and width $0.75 E_0$ for even filling factors
$\nu = 0, 2$. 
The difference occurs because excitations which are mixed by the direct interaction, 
such as the pair in Fig.~\ref{fig-diag} with $J_z=0$, partake in the CE for odd filling factors, but not for even ones (see Appendix \ref{app-se}).  Upon introducing an impurity, discrete localized CEs with energies $E$ appear both below and \emph{above} the band of continuum states. We plot the dimensionless energy $\mathcal{E}$ as a function of the dimensionless impurity coupling constant $g$:
\begin{eqnarray}
\label{eq-ener}
  \mathcal{E} & = & \frac{E}{E_0} \,  ,    \\
\label{eq-impcoupconst}
         g & = & \frac{V}{E_0 \ell_B^2}=\frac{\sqrt{3}}{2}\frac{W}{E_0} \left( \frac{a}{\ell_B} \right)^2 \sim W\sqrt{B}   \, .
\end{eqnarray}
Thus the impurity strength $W$ and magnetic field $B$ may be simultaneously tuned so that the quantity $g$ is left unchanged. Note that in this case the dimensionless energies $\mathcal{E}$ remain unchanged, indicating a scaling property of our theory.

Note that all the plotted energies have been renormalized: 
The lower continuum edge corresponds to the observable lowest cyclotron mode, 
$\hbar\tilde{\omega}_c = \hbar\omega_c + \delta\hbar\omega_c$, which has been set to the zero energy reference level. Here 
$\delta\hbar\omega_c = \Delta_{0 \uparrow \Uparrow}^{1 \uparrow \Uparrow}+\Gamma_{0 \uparrow \Uparrow}^{1 \uparrow \Uparrow}\simeq 0.7 E_0$ 
is the $e$-$e$ correction to the bare cyclotron energy, with $\Delta_{0 \uparrow \Uparrow}^{1 \uparrow \Uparrow}\approx 1.43E_0$ 
the total self energy correction and $\Gamma_ {0 \uparrow \Uparrow}^{1 \uparrow \Uparrow}=-0.75E_0$ the ``excitonic'' or vertex correction (see Appendix \ref{app-se} 
and Ref.~\onlinecite{FisDR09a}).
The size of square data points in Figs.~\ref{fig-Amz1nu1}--\ref{fig-Amz-1nu1} is proportional to the optical dipole transition matrix element 
squared $|d|^2$, see Sec.~\ref{subsec-op} above.
Notice that the majority of states are optically active, although some (marked as ``faint'') are considerably weaker than others, 
by at least three orders of magnitude. Some branches become brighter as they approach the band, whereas for others, their oscillator strength decreases. The physical reason for this behavior is at present unclear.

The dark states shown in Fig.~\ref{fig-Amz1nu1} have a special character. Out of the six possible types of excitation in our basis (see Fig.~\ref{fig-diag}), only the two excitations, 
$|{0 \! \downarrow \Downarrow} \;  { -1 \! \downarrow \Downarrow} \; J_z =1 \rangle$ and 
$|{0\! \uparrow \Downarrow} \; {-1 \! \uparrow \Downarrow } \; J_z=1\rangle$ contribute; see Eq.\ (\ref{Q}) for the definition of $|\mathcal{ N}_1  \mathcal{N}_2 J_z \rangle$. 
Having no amplitude on $|{1 \!\!\uparrow \Downarrow} \; { 0 \!\!\uparrow \Downarrow} \;  J_z=1\rangle$,
the only excitation out of the six which is optically active in $\sigma^+$, explains why they are dark. 
Explicitly they are created by the operator
\begin{equation}
\label{eq-rdag}
D^\dagger=\sum_m A(m)\left(c^\dagger_{0 \uparrow \Downarrow m} d^\dagger_{-1 \uparrow \Downarrow 2+m}-c^\dagger_{0 \downarrow \Downarrow m}d^\dagger_{-1 \downarrow \Downarrow 2+m}\right),
\end{equation}
with certain amplitudes $A(m)$ rapidly decreasing with $m$ [distance from impurity, Eq.~(\ref{dist})]. 
Although generally the total spin quantum number is not well defined for CEs, 
excitations (\ref{eq-rdag}) have a higher symmetry and are in fact spin triplet states $S=1, S_z=0$. 
Besides, the type of excitations is unusual in the following sense:  
All other excitations shown in Fig.~\ref{fig-Amz1nu1} have contributions from all six of the excitations which can be mixed. 
We note that the excitations with pseudospin flips are in general strongly mixed by the $\delta$-potential to those with no pseudospin flips.

The results indicate critical $g$ values for the formation of localized excitations, although within our approach it is difficult 
to resolve what may be a barely bound low-energy state from the lower continuum edge. The critical value for which bound states appear varies slightly depending on the filling factor $\nu$, light polarization $\sigma^{\pm}$ and the sublattice position $A,B$ of the impurity. 
For example, for a system with filling factor $\nu=-1$ illuminated by $\sigma^+$ circularly polarized light with an impurity on an $A$ sublattice 
[such a system will henceforth be denoted by ($A$, $\sigma^+$, $\nu=-1$)], there are no bound states for $-2.3<g<7.4$. If we then take $B=15$\,T and $\varepsilon=5$ for example, this corresponds to an impurity strength of approximately $-100\mathrm{\,eV}<W<340\mathrm{\,eV}$. Notice that short range potentials with rather large amplitudes correspond to vacancies.\cite{Bas08} It takes a very strong $\delta$-function impurity potential to localize excitations, since the impurity only couples to basis states where one of the particles has $m=0,1$. Strictly speaking, such energies are larger than the width of the $\pi$ band, where the electrons can be reasonably treated as massless Dirac fermions.  For such high values of the impurity potential, our results have only a qualitative nature.

In all cases, larger magnitudes of $g$ are required to push the states \emph{above} the band than to pull them below. 
This is because the states below the band form when either the electron or hole is nearer to the impurity and attracted to it; in this case the additional $e$-$h$ attraction also lowers the energy. States above the band form when one particle is held nearer to the impurity due to magnetic confinement, but is repelled by it, whilst the other particle is further away. In this case the $e$-$h$ attraction works against this, trying to lower the energy, so that a larger impurity strength is required to overcome this. 

The bound states move further from the band as the impurity coupling constant $g$ increases, as expected. 
An interesting question is what happens to the bound states when they approach the band. 
One possibility is they cease to exist, merging with the two-particle $e$-$h$ continuum.
The alternative is that some continue to exist within the band as quasibound states (resonances).
As a general rule,
the latter states have high probability amplitudes on the impurity and 
long-range oscillating tails, which make them non-normalizable.\cite{BazZP69}  
The existence of resonances is facilitated by the confining effect of the magnetic field.
Although we detect possible signatures of such resonant states, our method is not accurate enough to claim their existence. 
Interestingly, in all cases, the number of branches below the band for positive (negative) 
$g$ equals that of branches above the band for negative (positive) $g$.

In most cases for large enough $|g|$ values, bound states may be found both above and below the band for both $g>0$ and $g<0$. 
One exception (see Fig.~\ref{fig-Amz1nu1}) is the ($A$, $\sigma^+$, $\nu=-1$) system. 
In this case the impurity selection rule
forbids the hole to interact with the impurity, 
so that all diagonal matrix elements have the same sign as $g$. 
As a result, the states above the band are only seen for large enough positive $g$ and states below the band are only for large enough negative $g$. 
This behavior is mimicked by the ``conjugate''-symmetric (see Sec.~\ref{subsec-sym} below) system ($B$, $\sigma^-$, $\nu=1$)  except that the electron is forbidden to interact with the impurity. As a result, states above the band only appear for large enough negative values $g < 0$ and states below the band only for large enough positive $g > 0$ values. The other exception is the pair of ``sublattice'' symmetric systems 
($A, \sigma^-, \nu=2$) and ($B, \sigma^-, \nu=2$), where in both cases only the hole interacts with the impurity so that states above (below) the band only exist for large enough negative (positive) values of $g$. These are in turn ``conjugate''-symmetric to the systems ($B, \sigma^+, \nu=-2$) and ($A, \sigma^+, \nu=-2$) respectively.
The number of branches due to the electron-impurity and the number of branches due to the hole-impurity interaction can be correctly predicted by studying the transitions involved and the impurity selection rules. 

\subsection{Symmetries of the spectra}
\label{subsec-sym}

The above results indicate that the CEs possess a high level of symmetry. Moving the impurity from one sublattice to the other, 
$A \leftrightarrow B$, yields the same energies and oscillator strengths for filling factors $\nu=0,2$ (see Fig.~\ref{fig-Amz1nu2}). 
However, the results are qualitatively different
for filling factors $\nu=\pm1$ (compare Figs.~\ref{fig-Amz1nu1} and \ref{fig-Bmz1nu1}). 
This is because for odd filling factors such as $\nu=\pm1$, the valleys and thus sublattices are inequivalent.
This follows from the phenomenologically introduced valley splitting energy $\hbar\omega_{v}\tau_z$ [see Eq.~(\ref{eq-en})],
which leads to unequal occupancies of the two valleys. However, 
for even filling factors $\nu=0,2,\ldots$, the occupancies are the same and effectively there is no valley splitting. 
For the cases when the sublattices may be considered equivalent, the $A \leftrightarrow B$ 
correspondence of the energies and oscillator 
strengths is understood formally by observing that the Hamiltonians $\hat{H}_A$ and $\hat{H}_B$ are connected by a unitary transformation
$\hat{H}_B=\hat{U}\hat{H}_A\hat{U}^\dagger$. This transformation interchanges both valley and sublattice indices and is given by
\begin{equation}
\label{eq-un}
\hat{U}=
\left(
\begin{array}{cccc}
0 & 0 & 0 & \alpha \\
0 & 0 &\alpha  & 0 \\
0 & -\alpha^\ast & 0 & 0 \\
-\alpha^\ast & 0 & 0 & 0 \\
\end{array}
\right)  \, ,
\end{equation}
with $\alpha=e^{2\pi i/3}$.

Upon moving the impurity between the sublattices $A \leftrightarrow B$, symmetry may be restored also for odd filling factors. This requires the valleys to be interchanged, which can be achieved by reflecting about $n=0$. This entails flipping the spin and pseudospin quantum numbers and transforming electrons into holes and vice versa ($n \to -n$), which means that $\nu \leftrightarrow -\nu$ and the sign of the impurity potential and also the light polarization should be changed. Note that the LL structure in graphene, 
as well as its upper-lower cone dispersion in the absence of a magnetic field, is electron-hole symmetric. As a result, the energies and oscillator strengths of CEs induced by circularly polarized light in a system with filling factor $\nu$ and an impurity on the $A$ sublattice of strength $V$, correspond exactly to those of excitations induced by light circularly polarized in the opposite direction in a system with filling factor $-\nu$ and an impurity on the $B$ sublattice of strength $-V$. This symmetry may be expressed symbolically as 
\begin{equation}
   \label{symmetry}
     A, \nu, V, \sigma^\pm \longleftrightarrow B, -\nu, -V, \sigma^\mp \, .
\end{equation}
An example is shown in Fig.~\ref{fig-Amz1nu1}.
The above symmetry can be seen formally as follows. Switching the filling factor $\nu \to -\nu$ 
changes a pair 
excitation\footnote{A coherent superposition of such pair 
excitations constitutes collective excitations $|\mathcal{ N}_1  \mathcal{N}_2 J_z \rangle $,
see Eqs.~(\ref{eq-Qdag}) and (\ref{Q}).}
%
${|\mathcal{N}_1 m_1, \mathcal{N}_2 m_2 ; \nu \rangle \equiv  c^{\dag}_{\mathcal{N}_1 m_1} d^{\dag}_{\mathcal{N}_2 m_2}|\nu\rangle}$
according to
\begin{equation}
\label{ket}
|\mathcal{N}_1 m_1,             \mathcal{N}_2 m_2  ;  \nu \rangle \to 
 |\widetilde{\mathcal{N}}_2 m_2, \widetilde{\mathcal{N}}_1 m_1 ; - \nu \rangle \, , 
\end{equation}
where $\mathcal{N} \equiv  \{n,s,\tau \}$ and the conjugate 
$\widetilde{\mathcal{N}} \equiv \{-n,\tilde{s},\tilde{\tau}\}$
with $\tilde{s},\tilde{\tau}$ representing flipped spin and pseudospin. From this it follows that $J_z \to -J_z$. 
From the $e$-$h$ symmetry,
\begin{eqnarray}
 \nonumber
   \langle -\nu; \widetilde{\mathcal{N}}_2' m_2' , \widetilde{\mathcal{N}}_1' m_1' |
   \hat{H}_\mathrm{int}
   |\widetilde{\mathcal{N}}_2 m_2 , \widetilde{\mathcal{N}}_1 m_1 ; -\nu  \rangle  =  \\
\label{equiv-Hm} 
   \mbox{}\hphantom{XX} =  \langle \nu ; \mathcal{N}_1' m_1' , \mathcal{N}_2' m_2' |  \hat{H}_\mathrm{int}
   | \mathcal{N}_1 m_1, \mathcal{N}_2 m_2; \nu \rangle^\ast \: . 
   \end{eqnarray}
Hence the two matrices have the same energy eigenvalues and conjugate eigenvectors
leading to the same values of the dipole transition matrix elements $|d|^2$.
\section{Conclusions}
\label{sec-conclusions}

We have considered binding of neutral collective excitations
on a short-range impurity in graphene in strong magnetic fields.
The considered excitations are formed when an electron
is promoted to one of the empty Landau levels leaving behind a hole in a lower filled level.
This kind of problem requires treatment of interparticle (electron-hole) Coulomb,
electron-impurity, and hole-impurity interactions all on equal footing.
The scheme developed in this paper treats the impurity in the continuous approximation,
can be applied to arbitrary integer LL filling factors $\nu$ 
and gives the energies and optical strengths of  excitations localized on the impurity in graphene.
The concrete results we have presented and discussed here are for the $n=0$ Landau level filling factors $\nu = -1, 0, 1, 2$.

A separate study\cite{FisDR09c} using the same approach has shown that the $\delta$-impurity localized states do not seem to evolve smoothly
as a function of filling factor $\nu$ in contrast to the case of the Coulomb impurity.\cite{FisDR09a} This is perhaps because for the Coulomb impurity there are always four transitions with the same $J_z$ value for any filling factor $\nu$, whereas for the short-range impurity, where the intervalley scattering is significant, changing $\nu$  may change (i) the number of transitions that are mixed and (ii) their $J_z$ values.
This apparently affects the number of branches of bound states that are formed. Another difference between the  $\delta$-function and Coulomb impurities is that there are degenerate states for the Coulomb impurity
but not for the $\delta$-function. This is because the Coulomb impurity is sublattice symmetric,
whereas the $\delta$-function impurity introduces an asymmetry between sublattices and hence valleys,
thus breaking the SU(4) symmetry. 
Notice also, that the intervalley excitations $\mathbf{K} \leftrightarrows \mathbf{K}'$ 
$(\Uparrow \leftrightarrows \Downarrow)$
are admixed by a short range potential 
to optically-active intravalley excitations opening a possibility to introduce pseudospin flips by dipole photons.

We believe that the application of the continuous description to an extremely short-range $\delta$-impurity
systematically underestimates the impurity effect. Indeed, only a few impurity interaction channels are open in a given LL in graphene
as discussed above. Therefore, our predictions for the energies of bound states are mostly of a qualitative nature. It would be interesting to compare these with results of the tight-binding approach incorporating \emph{both} the impurity and the Coulomb interactions; this may be a subject of a separate study. We wish to stress that, on the other hand, the symmetries of the excitonic states and the optical selection rules leading to dark and bright states established in this paper should remain valid in other schemes. We believe that these qualitative predictions can be tested well in experiments.

A companion approach to studying the effect of short-range impurities would be the study of finite-range impurities, e.g.,  those which arise due to a disorder potential of Gaussian shape. Then the tuning of the height and width of the Gaussian would allow a continuous study of the transition from finite- to zero-range. Furthermore, one might then be able to quantitatively describe impurities induced by the presence of a suitably chosen substrate. We emphasize that the theoretical approach outlined in this work is equally applicable to such Gaussian disorder.


\appendix
\section{Derivation of single particle Hamiltonian}
\label{app-ham}

The Hamiltonians, $\hat{H}_{\mu}$, for a single electron in the presence of a $\delta$-function impurity on a $\mu = A, B$ sublattice site at the origin, are derived below within the tight binding and continuum approximations. The approach follows that used in Ref.~\onlinecite{Sem84} to derive the free electron Hamiltonian, 
$\hat{H}_0$. The results are in agreement with the expressions obtained in Ref.~\onlinecite{AndN98} using a different approach. For an impurity at the origin and on the $A$ sublattice, the tight binding Hamiltonian is
\begin{equation}
\label{eq-tbhama}
 \hat{\mathcal{H}}_A = 
  t\sum_{\mathbf{R}_i, \bm{\tau}_j}\left( a^\dagger_{\mathbf{R}_i}b_{\mathbf{R}_i+\bm{\tau}_j} 
                                        + b^\dagger_{\mathbf{R}_i+\bm{\tau}_j}a_{\mathbf{R}_i} \right) 
                        + W a^\dagger_0 a_0 \, ,
\end{equation}
$\hat{\mathcal{H}}_A \equiv \hat{\mathcal{H}}_0 + \hat{\mathcal{V}}_A$. 
Here $t$ is the nearest neighbors hopping energy, $\mathbf{R}_i$ the real space lattice vectors, $\bm{\tau}_j$ the nearest neighbor vectors 
as indicated in Fig.~\ref{fig-lat} and $a^\dagger(b^\dagger)$ the operator which creates an electron on the $A$($B$) sublattice. 
The Fourier transforms of the annihilation operators are
\begin{equation}
a_{\mathbf{R}_l} = \frac{1}{\sqrt{N_\mathrm{c}}} \sum_{\mathbf{k}} \mathrm{exp} \left(i\mathbf{k} \cdot \mathbf{R}_l \right) a \left(\mathbf{k}\right),
 \label{eq-fta}
\end{equation}
\begin{equation}
b_{\mathbf{R}_l+\bm{\tau}_m}=\frac{1}{\sqrt{N_\mathrm{c}}} \sum_{\mathbf{k}} \mathrm{exp} \left[i\mathbf{k} \cdot \left( \mathbf{R}_l+\bm{\tau}_m\right) \right]b\left(\mathbf{k}\right),
 \label{eq-ftb}
\end{equation}
where $N_\mathrm{c}$ is the number of unit cells in the crystal.
Substituting these into Eq.~(\ref{eq-tbhama}) and expanding about the $\mathbf{K}$, $\mathbf{K'}$ minima (the continuum approximation) yields
\begin{eqnarray}
\label{eq-conapproxa}
 \hat{\mathcal{H}}_0 &= & t\sum_{\mathbf{q}, \bm{\tau}_j}
\Big[ 
    \Big( e^{i\left(\mathbf{K} + \mathbf{q} \right) \cdot \bm{\tau}_j} 
	        a^\dagger_{\mathbf{K}}\left(\mathbf{q}\right)b_{\mathbf{K}}\left(\mathbf{q} \right) 
			                    + \mathrm{H.c.} \Big) \\ \nonumber
%
& & \mbox{}\hphantom{X..} +
    \Big( e^{i\left(\mathbf{K}' + \mathbf{q} \right) \cdot \bm{\tau}_j} 
            a^\dagger_{\mathbf{K}'}\left(\mathbf{q}\right)b_{\mathbf{K}'}\left(\mathbf{q} \right) 
			                   + \mathrm{H.c.} \Big)  \Big]  \, , \\ 
\nonumber
\hat{\mathcal{V}}_A & = & \frac{W}{N_\mathrm{c}} \sum_{\mathbf{q},\mathbf{q}'}
\Big[
       a^\dagger_{\mathbf{K}}\left(\mathbf{q}\right)a_{\mathbf{K}}\left(\mathbf{q}' \right) 
	 + a^\dagger_{\mathbf{K}'}\left(\mathbf{q}\right)a_{\mathbf{K}'}\left(\mathbf{q}' \right) \\ 
\label{eq-conapproxAA}
& & \mbox{}\hphantom{XXX} + \Big( 
       a^\dagger_{\mathbf{K}}\left(\mathbf{q}\right)a_{\mathbf{K}'}\left(\mathbf{q}' \right) 
	  + \mathrm{H.c.} \Big)
\Big],
\end{eqnarray}
where, e.g., $a_{\mathbf{K}}\left(\mathbf{q}\right) \equiv a\left(\mathbf{K} + \mathbf{q}\right)$.
Since $|\mathbf{q}| \ll 1/a$, we can expand in (\ref{eq-conapproxa}) 
to linear order in $|\mathbf{q}|$. The summation over $\bm{\tau}_j$ is then performed using 
for graphene
$\sum_{j}e^{i \mathbf{K} \cdot \bm{\tau}_j} = \sum_{j}e^{i \mathbf{K}' \cdot \bm{\tau}_j}=0$ 
and 
$\sum_{j}e^{i \mathbf{K} \cdot \bm{\tau}_j}\bm{\tau}_j = -\frac{\sqrt{3}}{2}a \alpha^\ast \left(i,1\right)$, 
$\sum_{j}e^{i \mathbf{K}' \cdot \bm{\tau}_j}\bm{\tau}_j=\frac{\sqrt{3}}{2}a \left(i,-1\right)$. 
This gives
\begin{eqnarray}
\label{eq-smallqa}
 \hat{\mathcal{H}}_0 
& = &
\frac{\sqrt{3}}{2}at \sum_{\mathbf{q}}
  \Big[
         \Big( \alpha q_+ b^\dagger_{\mathbf{K}} \left(\mathbf{q}\right)a_{\mathbf{K}}\left(\mathbf{q} \right) + \mathrm{H.c.} \Big) \\
   & & \mbox{}\hphantom{XXXX}
       - \Big( q_+ a^\dagger_{\mathbf{K}'}\left(\mathbf{q}\right)b_{\mathbf{K}'}\left(\mathbf{q} \right) + \mathrm{H.c.} 
    \Big) 
  \Big] \nonumber \, , 
\end{eqnarray}
where $q_\pm=q_x\pm iq_y$ and $\alpha = e^{2\pi i/3}$.
To return to real space, we define the field operator $\Psi_{A \mathbf{K}}\left(\mathbf{r}\right)$ according to
\begin{equation}
\label{eq-ftreala}
a_\mathbf{K}\left(\mathbf{q}\right)=\int \! d\mathbf{r} \, 
   e^{-i \left(\mathbf{K} + \mathbf{q}\right) \cdot \mathbf{r} }\Psi_{A \mathbf{K}}\left(\mathbf{r}\right)
\end{equation}
and  $\Psi_{A \mathbf{K}'}\left(\mathbf{r}\right),\Psi_{B \mathbf{K}}\left(\mathbf{r}\right),\Psi_{B \mathbf{K}'}\left(\mathbf{r}\right)$ similarly. Substituting these expressions into Eqs.~(\ref{eq-conapproxAA}) and (\ref{eq-smallqa}) and converting the sum over $\mathbf{q}$ 
into an integral gives
\begin{widetext}
\begin{eqnarray}
\label{eq-finala0}
 \hat{\mathcal{H}}_0 & = & 
\hbar v_\mathrm{F}
      \int \! \frac{S\,d\mathbf{q}}{(2\pi)^2}  \int \! d\mathbf{r} 
                                           	   \int \! d\mathbf{r}' \, 
\Big[ \Big( \alpha q_+ e^{i \left( \mathbf{K} + \mathbf{q}\right) \cdot \left(\mathbf{r}-\mathbf{r}' \right) }
                  \Psi^\dagger_{B \mathbf{K}}\left(\mathbf{r}\right)
				  \Psi_{A \mathbf{K}}\left(\mathbf{r}'\right)
				  + \mathrm{H.c.} \Big)
\\ \nonumber
 && \mbox{}\hphantom{XXXXXXXXXXXX}  
  -  \Big(  q_+ e^{i \left( \mathbf{K}' + \mathbf{q}\right) \cdot \left(\mathbf{r}-\mathbf{r}' \right) }
                   \Psi^\dagger_{A \mathbf{K}'}\left(\mathbf{r}\right) \Psi_{B \mathbf{K}'}\left(\mathbf{r}'\right)
         + \mathrm{H.c.} \Big)	\Big]  \, ,
\\ \nonumber			   
\hat{\mathcal{V}}_A  & = &  \frac{W}{N_\mathrm{c}} 
            \int \! \frac{S\,d\mathbf{q}}{(2\pi)^2} 
			\int \! \frac{S\,d\mathbf{q}'}{(2\pi)^2} \,  
			\int \! d\mathbf{r} 
			\int \! d\mathbf{r}' \,
\Big[ \Big(
    e^{ i \left(\mathbf{K} + \mathbf{q} \right) \cdot \mathbf{r}  }
	e^{-i \left(\mathbf{K} + \mathbf{q}'\right) \cdot \mathbf{r}' }
        \Psi^\dagger_{A \mathbf{K}}\left(\mathbf{r}\right)\Psi_{A \mathbf{K}}\left(\mathbf{r}'\right)
+ \{ \mathbf{K} \rightarrow \mathbf{K}' \} \Big)  \\   
\label{eq-ftreal}
&& \mbox{}\hphantom{XXXXXXXXXXXXXXXX} + \Big( e^{ i \left(\mathbf{K} + \mathbf{q}\right) \cdot \mathbf{r} }
           e^{-i \left(\mathbf{K}' + \mathbf{q}'\right) \cdot \mathbf{r}' }
       \Psi^\dagger_{A \mathbf{K}}\left(\mathbf{r}\right)
	   \Psi_{A \mathbf{K}'}\left(\mathbf{r}'\right)  
	   + \{ \mathbf{K} \rightleftarrows \mathbf{K}' \} \Big)
\Big] \, ,
\end{eqnarray}
where $S$ is the area of the graphene sheet and $v_\mathrm{F}=\sqrt{3}at/2 \hbar$, the Fermi velocity. The momentum operators are introduced by observing 
$q_\pm e^{i\mathbf{q}\cdot \left(\mathbf{r}-\mathbf{r}'\right)}=\frac{1}{\hbar}p_\pm e^{i\mathbf{q}\cdot \left(\mathbf{r}-\mathbf{r}'\right)}$ and replacing in the continuous approximation the canonical momentum 
$\mathbf{p} = - i\hbar\bm{\nabla}$ by the kinematic momentum $\bm{\Pi}=\mathbf{p} + \frac{e}{c}\mathbf{A}$ 
for the case of a magnetic field. The result may be expressed as  
\begin{eqnarray}
\label{eq-mata}
 \hat{\mathcal{H}}_A &= &
  \int \! d\mathbf{r} \, 
\left(
\begin{array}{r}
\Psi^\dagger_{A \mathbf{K}}\left(\mathbf{r}\right)  \\
\alpha \Psi^\dagger_{B \mathbf{K}}\left(\mathbf{r}\right)  \\
\Psi^\dagger_{A \mathbf{K}'}\left(\mathbf{r}\right)  \\
 -\Psi^\dagger_{B \mathbf{K}'}\left(\mathbf{r} \right) 
\end{array}
\right)^{\!\!\!\mathrm{T}}
\left(
\begin{array}{cccc}
V \delta (\mathbf{r})  & v_\mathrm{F}\Pi_- &  V \delta (\mathbf{r}) & 0 \\
v_\mathrm{F}\Pi_+ & 0 & 0 &  0 \\
V \delta (\mathbf{r}) & 0 & V \delta (\mathbf{r}) &v_\mathrm{F} \Pi_+ \\
0 &  0  & v_\mathrm{F}\Pi_- & 0 
\end{array}
\right)
\left(
\begin{array}{r}
\Psi_{A \mathbf{K}}\left(\mathbf{r}\right) \\
\alpha^\ast \Psi_{B \mathbf{K}}\left(\mathbf{r}\right) \\
\Psi_{A \mathbf{K}'}\left(\mathbf{r}\right) \\
-\Psi_{B \mathbf{K}'}\left(\mathbf{r}\right)
\end{array}
\right) \, ,
\end{eqnarray}
which gives Eqs.~(\ref{eq-ham_free}) and (\ref{eq-va}). In Eq.\ (\ref{eq-mata}), $V=\sqrt{3}Wa^2/2$ and the superscript $\mathrm{T}$ denotes transposition.
Analogous derivation may be repeated for the case when the impurity is on the $B$ sublattice chosen to be \emph{at the origin} again, 
so the nearest neighbor vectors are given by the $\bm{\delta}_j$ indicated in Fig.~\ref{fig-lat}, whilst the valley definitions remain the same. This yields [cf.~Eq.~(\ref{eq-vb})]
\begin{eqnarray}
\label{eq-matb}
 \hat{\mathcal{H}}_B &= &
 \int \! d\mathbf{r} \,
\left(
\begin{array}{r}
       \Phi^\dagger_{A \mathbf{K}}\left(\mathbf{r}\right)   \\
\alpha \Phi^\dagger_{B \mathbf{K}}\left(\mathbf{r}\right)  \\
       \Phi^\dagger_{A \mathbf{K}'}\left(\mathbf{r}\right)  \\
      -\Phi^\dagger_{B \mathbf{K}'}\left(\mathbf{r}\right) 
\end{array}
\right)^{\!\!\!\mathrm{T}}
\left(
\begin{array}{cccc}
 0 & v_\mathrm{F}\Pi_- & 0 & 0 \\
v_\mathrm{F}\Pi_+ & V \delta (\mathbf{r}) & 0 & -\alpha^\ast V \delta (\mathbf{r}) \\
0 & 0 & 0 & v_\mathrm{F}\Pi_+ \\
0 & -\alpha V \delta (\mathbf{r})  & v_\mathrm{F}\Pi_- & V \delta (\mathbf{r}) 
\end{array}
\right)
\left(
\begin{array}{r}
\Phi_{A \mathbf{K}}\left(\mathbf{r}\right) \\
\alpha^\ast \Phi_{B \mathbf{K}}\left(\mathbf{r}\right) \\
\Phi_{A \mathbf{K}'}\left(\mathbf{r}\right) \\
-\Phi_{B \mathbf{K}'}\left(\mathbf{r}\right)
\end{array}
\right) \, .
\end{eqnarray}
\end{widetext}
Note that different bases are used in Eqs.~(\ref{eq-mata}) and (\ref{eq-matb}) due to the translation; 
they are related through $\Psi\left(\mathbf{r}\right)=\Phi\left(\mathbf{r}-\bm{\tau}_j\right)$.

\section{Higher order terms}
\label{app-high}
Here we estimate the relative magnitudes of higher-order energy corrections coming from the interaction with the $\delta$-impurity.
 Suppose we have an electrically neutral system so that the lower cone is completely filled and the upper cone is completely empty. Let's denote a single electron state by $|n s \tau m\rangle=c^{\dag}_{n s \tau m} |0 \rangle$. As an example, we will suppose the impurity is on the $B$ sublattice and calculate the ratio of the first and second order corrections to the single particle energy of an electron in the $n=0$ LL with, say, spin, pseudospin equal to $\downarrow, \Uparrow $. As indicated by the form of the impurity matrix element in Eq.~(\ref{eq-impme}), 
an electron in the $n=0$ LL will only interact with the impurity if $m=0$. 
Hence we consider the state $|0 \! \downarrow \Uparrow \! 0\rangle$. The results are qualitatively the same for other scenarios. The first order correction is 
$E^{(1)}={\mathcal{V}_B}_{0  \downarrow \Uparrow 0}^{0 \downarrow \Uparrow 0} = V/2 \pi \ell_B^2 \sim B.$ 
The second order correction is 
\begin{equation}
\label{appBe2}
E^{(2)}={\sum_{n=1}^{n_\mathrm{c}} \sum_{m=0}^\infty}\dfrac{|{\mathcal{V}_B}_{0 \downarrow \Uparrow 0}^{n \downarrow \Uparrow m}|^2
+|{\mathcal{V}_B}_{0 \downarrow \Uparrow 0}^{n \downarrow \Downarrow m} |^2}{\epsilon_{0\downarrow\Uparrow}-\epsilon_{n\downarrow \Uparrow}}\,.
\end{equation}
This represents virtual processes where an electron spontaneously hops to a higher LL and then decays back to its original state. The sum over the LL index $n$ is cut off at $n_\mathrm{c}$, which corresponds to the energy limit of where the linear dispersion relation ceases to hold. 
Using ${\mathcal{V}_B}_{0  \downarrow \Uparrow 0}^{n \downarrow \Uparrow m}=\delta_{n, m}\frac{(-i)^m V}{2 \sqrt{2} \pi \ell_B^2}$ 
and ${\mathcal{V}_B}_{0 \downarrow \Uparrow 0}^{ n \downarrow \Downarrow m}=-\alpha\delta_{n-1, m}\frac{(-i)^m V}{2 \sqrt{2} \pi \ell_B^2}$, 
we obtain for Eq.~\eqref{appBe2}
\begin{equation}
\label{appB_e2}
E^{(2)}=-\dfrac{V^2}{4 \sqrt{2} \pi^2 \hbar v_\mathrm{F} \ell_B^3}\sum_{n=1}^{n_\mathrm{c}} \dfrac{1}{\sqrt{n}}\,.
\end{equation}
This gives the ratio of the first and second order terms as
\begin{equation}
 \label{appBratio}
\dfrac{E^{(2)}}{E^{(1)}}=-\dfrac{V}{2 \sqrt{2} \pi \hbar v_\mathrm{F} \ell_B}\sum_{n=1}^ {n_\mathrm{c}}\dfrac{1}{\sqrt{n}}\,.
\end{equation}
It can be shown that the ratio for the $M^{\mathrm{th}}$ order correction behaves as 
$\frac{E^{(M)}}{E^{(1)}}\sim \left( \frac{V}{\hbar v_\mathrm{F} \ell_B}\right)^{M-1} \sim V^{M-1} B^{(M-1)/2}$ 
so that the perturbation theory is applicable when the dimensionless parameter 
$V/\hbar v_\mathrm{F} \ell_B  \ll 1$. 
Note that in graphene this parameter increases with magnetic field $B$ 
making the formal limit $B \rightarrow \infty$ singular; 
this is unlike the situation in the 2DEG, where the corresponding parameter is independent of $B$.
In order to obtain a numerical value for Eq.~\eqref{appBratio}, 
we need to estimate $n_c$. 
Since the linear dispersion relation holds provided $|\mathbf{q}|\ll 1/a$, $n_c$ is defined according to 
$\hbar \omega_c \sqrt{n_c}=\hbar v_\mathrm{F}/a$. This yields $n_c\approx 360$ for $B=15$\,T. 
Using these parameters and taking $W=1$\,eV for the impurity potential, 
we estimate $ |E^{(2)}| / |E^{(1)}| \approx 0.05$. 
This shows that second and higher order terms in the perturbation expansion can be ignored for not too 
strong impurities and not in too strong fields $B$.

\section{Self-energy corrections}
\label{app-se}

Kohn's theorem\cite{Koh61} states that for electron systems with parabolic
dispersion
$E(p)=p^2/2m$, the observed cyclotron resonance energy is independent of the $e$-$e$
interactions
and is equal to the bare non-interacting value $\hbar\omega_c = \hbar
eB/mc$.
The theorem does not hold in graphene with massless electrons.
Formally, in conventional systems with a constant charge-to-mass ratio,
irrespective of dimensionality (2D or 3D),
(i) the Coulomb vertex correction $\Gamma<0$ to the energy [the $e$-$h$ ``exciton'' lowering of energy analogous to Eq.\ (\ref{eq-w})] is precisely
compensated by (ii) the positive difference between the self-energy
corrections $\Delta>0$ to the final electron and initial hole states;
the latter are due to the exchange with electrons in filled states.
In graphene, the single particle energies also have self energy corrections, since each
electron (hole)
interacts via the exchange attractive (repulsive) interaction with all other
electrons in the Dirac sea with same spin and pseudospin. However, this difference is not
compensated by the excitonic effects
in graphene, $\Delta+\Gamma>0$, leading to the Coulomb renormalization of the bare cyclotron resonance energy
$\hbar\omega_c \rightarrow  \hbar\tilde{\omega}_c =\hbar\omega_c +
\delta\hbar\omega_c$ with
$\delta\hbar\omega_c =\Delta+\Gamma\approx 0.7 E_0$,
which may be absorbed in the renormalization of the Fermi velocity.\cite{JiaHTW07,IyeWFB07}

\begin{figure}[t]
  \centering
 \includegraphics[width=0.37\textwidth]{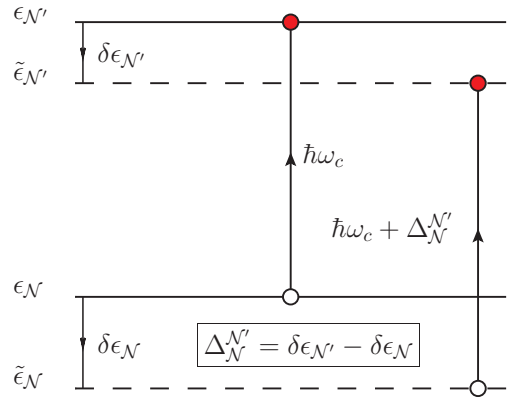}
  \caption {(Color online) Diagram illustrating the self-energy corrections, which contribute to the cyclotron resonance
renormalization (increase). Note that ``excitonic'' corrections $\Gamma$ to the cyclotron resonance energy 
due to the Coulomb vertex are not shown here.
}
\label{fig-selfeng}
\end{figure}

Here we provide some details about the self-energy Coulomb correction to
single-particle energies in graphene. The situation is represented schematically in Fig.~\ref{fig-selfeng}.
Such corrections are negative, of the order of $E_0$
and decrease in magnitude as the LL number of the state in question
increases.
Specifically,
the exchange interaction energy for an electron in state
$\mathcal{N}_1m_1$ and
a hole in state $\mathcal{N}_2 m_2$ is given by the matrix element
$\mathcal{U}_{\mathcal{N}_1 m_1   \mathcal{N}_2 m_2}^{\mathcal{N}_2 m_2
\mathcal{N}_1 m_1}$, as seen in Eq.~(\ref{eq-w}).
Analogously, the interaction energy due to exchange between
two electrons in states, $\mathcal{N} m$ and $\mathcal{N}'m'$ is given by
$-\mathcal{U}_{\mathcal{N} m   \mathcal{N}' m'}^{\mathcal{N}' m'\mathcal{N} m}$, 
where the sign change is due to switching $h \rightarrow e$. 
Hence the correction to the energy of an
electron in the LL $n$ with spin $s$, pseudospin $\tau$ and oscillator quantum number $m$ 
due to exchange with electrons in the LL $n'$, 
which has a fully filled same $s,\tau$ sublevel, is
\begin{equation}
\label{eq-se}
 \delta\epsilon\left( n,n'\right)=-\frac{1}{2}\sum_{m'}
   \mathcal{U}_{n \tau s  m\hspace{2mm}   n'\tau s m'}^{n' \tau s m' \hspace{1mm}   n \tau s m} \, ,
\end{equation}
which may be written in terms of 2DEG matrix elements.\cite{FisDR09a}
We remark that this expression is independent of the values of $m$, $s$ and $\tau$.
The complete self-energy correction for an electron in
the LL $n$ is given by
$\delta\epsilon_\mathcal{N} = \sum_{n'} \delta\epsilon\left(n,n'\right)$, where the range over which $n'$ is summed depends on the filling factor and the values of $s$, $\tau$.
The renormalised single particle energies in Eq.~(\ref{eq-bigham}) are then given by 
$\tilde{\epsilon}_\mathcal{N}=\epsilon_\mathcal{N}+\delta\epsilon_\mathcal{N}$.

As a means of illustration, we shall give the self-energy corrections
to the energies of those excitations mixed for the ($A$, $\sigma^+$, $\nu=-1$) system, shown in bold (red) in Fig.~\ref{fig-diag}.
Let $\Delta_{\mathcal{N}}^{\mathcal{N}'}$ denote the self energy correction to the bare cyclotron resonance for an excitation with a hole in state $\mathcal{N}$ and an electron in state $\mathcal{N}'$. The four excitations with no spin or pseudospin flip have the same
corrections. Specifically for $s \tau$ = $\uparrow \Downarrow$,$\downarrow \Uparrow$,$\downarrow \Downarrow$
\begin{eqnarray}
\label{eq-se2a}
 \Delta_{0 \uparrow \Uparrow}^{1 \uparrow \Uparrow} & = &\Delta_{-1 s \tau}^{\hphantom{b.} 0 s \tau} \, ,  \\
\label{eq-se2}
\Delta_{-1 s \tau}^{\hphantom{b.} 0 s \tau}& \equiv & \delta\epsilon_{0 s \tau}-\delta\epsilon_{-1 s \tau}\\ \nonumber
&= &  \sum_{n=1}^{n_c}\left[\delta\epsilon\left(
0,-n\right)-\delta\epsilon\left( -1,-n\right) \right] \\ \nonumber
&= &
\frac{E_0}{8\sqrt{\pi}}\sum_{n=1}^{n_c}\frac{\Gamma\left(n+\frac{1}{2}\right)}{n!}\frac{4\sqrt{n}+3}{2n-1} \ ,
\end{eqnarray}
where $\Gamma$ is the $\Gamma$-function and $n_c$ is the cutoff value for the LL index, as described in Appendix
\ref{app-high}.
One may show that $\Delta_{0 \uparrow \Uparrow}^{1 \uparrow \Uparrow} \sim
\mathrm{ln}\, n_c$,\cite{IyeWFB07}
so that for large enough values of $n_c$, the result hardly changes for small
alterations in $n_c$.
For $B=15$\,T, $n_c\approx360$ as before and $\Delta_{0 \uparrow \Uparrow}^{1 \uparrow
\Uparrow}\approx 1.43E_0$.
\footnote{This value differs slightly from the one we gave in Ref.~\onlinecite{FisDR09a}. There the self energy correction due to exchange interactions with the lower cone only was given as $0.92E_0$. This corresponds to a total self energy correction, including exchange with both the lower and upper cones, of $\Delta(n_c=1870)=0.75E_0+0.92E_0=1.67E_0$. It is this value which should be compared with $\Delta(n_c=360)=1.43E_0$ quoted above. The remaining difference is because in this paper we use $n_c=360$, whereas in Ref.~\onlinecite{FisDR09a} we took $n_c=1870$ as used in Ref.~\onlinecite{IyeWFB07}.}
For the two excitations with a pseudospin flip,
$\Delta_{0 \uparrow \Uparrow}^{1 \uparrow \Downarrow}  =
 \Delta_{-1 \uparrow \Uparrow}^{ \hphantom{b.} 0 \uparrow \Downarrow} =
 \Delta_{0 \uparrow \Uparrow}^{1 \uparrow \Uparrow}+0.25E_0$.
This difference in self-energy corrections means that for excitations
containing i.\ excitons with no spin/pseudospin flip
and ii.\ pairs of excitons mixed by the direct interaction,
the band width is greater by $0.25E_0$,
than for excitations containing only one of these types of excitons.
This leads to different band widths for different filling factors $\nu =0,2$ and $\nu=-1,1$ (see beginning of Sec.~\ref{subsec-en}).
\begin{acknowledgments}
We thank V.\;I.\ Fal'ko for illuminating discussions.
ABD gratefully acknowledges funding from the Cottrell Research Corporation (CSUB).
\end{acknowledgments}

\begin{thebibliography}{10}

\bibitem{Wal47}
P.~R. Wallace, Phys. Rev. {\bf 71},  622  (1947).

\bibitem{Sem84}
G.~W. Semenoff, Phys. Rev. Lett. {\bf 53},  2449  (1984).

\bibitem{CasGPN09}
{A. H. Castro Neto}, F. Guinea, N.~M.~R. Peres, K.~S. Novoselov, and A.~K.
  Geim, Rev. Mod. Phys. {\bf 81},  109  (2009).

\bibitem{AbeABZ10}
D.~S.~L. Abergel, V. Apalkov, J. Berashevich, K. Ziegler, and T. Chakraborty,
  Advances in Physics {\bf 59},  261  (2010).

\bibitem{KosVSL11}
S.~O. Koswatta, A. Valdes-Garcia, M.~B. Steiner, Y.-M. Lin, and P. Avouris,
  (2011), {ArXiv}:1105.1060.

\bibitem{NovGMJ04}
K.~S. Novoselov, A.~K. Geim, S.~V. Morozov, D. Jiang, Y. Zhang, S.~V. Dubonos,
  I.~V. Grigorieva, and A.~A. Firsov, Science {\bf 306},  666  (2004).

\bibitem{KotUPC10}
V.~N. Kotov, B. Uchoa, V.~M. Pereira, {A. H. Castro Neto}, and F. Guinea,
  (2010).

\bibitem{NomM06}
K. Nomura and A.~H. MacDonald, Phys. Rev. Lett. {\bf 96},  256602  (2006).

\bibitem{NovGMJ05}
K.~S. Novoselov, A.~K. Geim, S.~V. Morozov, D. Jiang, M.~I. Katsnelson, I.~V.
  Grigorieva, S.~V. Dubonos, and A.~A. Firsov, Nature {\bf 438},  197  (2005).

\bibitem{ZhaTSK05}
Y. Zhang, Y.-W. Tan, H.~L. Stormer, and P. Kim, Nature {\bf 438},  201  (2005).

\bibitem{DuSDL09}
X. Du, I. Skachko, F. Duerr, A. Lucian, and E.~Y. Andrei, Nature {\bf 462},
  192  (2009).

\bibitem{BolGSS09}
K.~I. Bolotin, F. Ghahari, M.~D. Shulman, H.~L. Stormer, and P. Kim, Nature
  {\bf 462},  196  (2009).

\bibitem{IyeWFB07}
A. Iyengar, J. Wang, H.~A. Fertig, and L. Brey, Phys. Rev. B {\bf 75},  125430
  (2007).

\bibitem{BycM08}
{Yu. A. Bychkov} and G. Martinez, Phys. Rev. B {\bf 77},  125417  (2008).

\bibitem{RolFG10b}
R. Roldan, J.-N. Fuchs, and M.~O. Goerbig, Phys. Rev. B {\bf 82},  205418
  (2010).

\bibitem{TahS08}
M. Tahir and K. Sabeeh, J. Phys.: Condens. Matter {\bf 20},  425202  (2008).

\bibitem{Shi09}
K. Shizuya, Phys. Rev. B {\bf 79},  165402  (2009).

\bibitem{FisRD10}
{A. M. Fischer}, {R. A. R\"omer}, and {A. B. Dzyubenko}, EPL {\bf 92},  37003
  (2010).

\bibitem{PerGLP06}
V.~M. Pereira, F. Guinea, J.~M.~B. Lopes~dos Santos, N.~M.~R. Peres, and A.~H.
  Castro~Neto, Phys. Rev. Lett. {\bf 96},  036801  (2006).

\bibitem{Bas08}
D.~M. Basko, Phys. Rev. B {\bf 78},  115432  (2008).

\bibitem{FisDR09a}
A.~M. Fischer, A.~B. Dzyubenko, and R.~A. R\"omer, Phys. Rev. B {\bf 80},
  165410  (2009).

\bibitem{AndN98}
T. Ando and T. Nakanishi, J. Phys. Soc. Japan {\bf 67},  1704  (1998).

\bibitem{DivM84}
D.~P. DiVincenzo and E.~J. Mele, Phys. Rev. B {\bf 29},  1685  (1984).

\bibitem{GoeMD06}
M.~O. Goerbig, R. Moessner, and B. Dou\c{c}ot, Phys. Rev. B {\bf 74},  161407
  (2006).

\bibitem{FisRD11}
A.~M. Fischer, R.~A. R\"omer, and A.~B. Dzyubenko, J.\ Phys.:\ Conf.\ Ser. {\bf
  286},  012054  (2011).

\bibitem{GusSC07}
V.~P. Gusynin, S.~G. Sharapov, and J.~P. Carbotte, Phys. Rev. Lett. {\bf 98},
  157402  (2007).

\bibitem{AbeF07}
D.~S.~L. Abergel and V.~I. Fal'ko, Phys. Rev. B {\bf 75},  155430  (2007).

\bibitem{Dzy90}
A.~B. Dzyubenko, Solid State Commun. {\bf 74},  409  (1990).

\bibitem{BazZP69}
{\em Scattering, Reactions and Decay in Non-Relativistic Quantum Mechanics},
  edited by A.~I. Baz', Y.~B. Zel'dovich, and A.~M. Perelomov (Israel Program
  for Scientific Translations, Jerusalem, 1969).

\bibitem{FisDR09c}
A.~M. Fischer, A.~B. Dzyubenko, and R.~A. R\"omer, {Ann. Phys. (Leipzig)} {\bf
  18},  944  (2009).

\bibitem{Koh61}
W. Kohn, Phys. Rev. {\bf 123},  1242  (1961).

\bibitem{JiaHTW07}
Z. Jiang, E.~A. Henriksen, L.~C. Tung, Y.~J. Wang, M.~E. Schwartz, M.~Y. Han,
  P.~P. Kim, and H.~L. Stormer, Phys. Rev. Lett. {\bf 98},  197403  (2007).

\end{thebibliography}

\end{document}